\documentclass[12pt]{article}
\usepackage{amssymb,amsfonts,amsmath,epsfig}

\renewcommand{\theequation}{\arabic{section}.\arabic{equation}}
\newcommand{\gapprox}{%
\mathrel{%
\setbox0=\hbox{$>$}\raise0.6ex\copy0\kern-\wd0\lower0.65ex\hbox{$\sim$}}}
\newcommand{\SU}{\mathop{\rm SU}}
\newcommand{\SO}{\mathop{\rm SO}}
\newcommand{\be}{\begin{equation}}
\newcommand{\ee}{\end{equation}}
\newcommand{\bea}{\begin{eqnarray}}
\newcommand{\eea}{\end{eqnarray}}
\def\CN{$\mathcal{N}$}
\def\CH{$\mathcal{H}$}
\newcommand{\bref}[1]{(\ref{#1})}

\begin{document}

\begin{titlepage}

\setcounter{page}{0}
\begin{flushright}
Imperial/TP/2-03 /7 \\
\end{flushright}

\vspace{5mm}
\begin{center}
{\Large {\bf Semi-classical string solutions for \CN=1 SYM}}
\vspace{10mm}

{\large
Josep M. Pons$^{a,b}$ and Pere Talavera$^{c}$} \\
\vspace{5mm}
$^a${\em
Departament d'Estructura i Constituents de la Mat\`eria,
Universitat de Barcelona,\\
Diagonal 647, E-08028 Barcelona, Spain.}\\[.1cm]
$^b${\em Theoretical Physics Group,
Blackett Laboratory, Imperial College London, SW7 2BZ, U.K.}\\[.1cm]
$^c${\em
Departament de F{\'\i}sica i Enginyeria Nuclear,
Universitat Polit\`ecnica de Catalunya,\\
Jordi Girona 31, E-08034 Barcelona, Spain.}
\vspace{5mm}

\vspace{20mm}
\centerline{{\bf{Abstract}}}
\end{center}
\vspace{5mm}

We study semi-classically the dynamics of string solitons in the
Maldacena-Nu\~nez background, dual in the infra-red to \CN=1,
$d=4$ SYM. For closed string configurations rotating in the $S^2
\times \mathbb{R}$ space wrapped by the stack of $N$ D-branes we
find a behaviour that indicates the decoupling of the stringy
Kaluza-Klein modes with sufficiently large R-symmetry charge. We
show that the spectrum of a pulsating string configuration in
$S^2$ coincides with that of a N=2 super Sine-Gordon model. Closed
string configurations spinning in the transversal $S^3$ give a
relation of the energy and the conserved angular momentum
identical to that obtained for configurations spinning in the
$S^5$ of the $AdS_5\times S^5$, dual to \CN =4 SYM. In order to
obtain non-trivial relations between the energy and the spin, we
also consider conical-like configurations stretching along a
radial variable in the unwrapped directions of the system of
D-branes and simultaneously along the transversal direction. We
find that in this precise case, these configurations are unstable
--contrary to other backgrounds, where we show that they are
stable. We point out that in the Poincar\'e-like coordinates used
for the Maldacena-Nu\~nez background it seems that it is not
possible to reproduce the well-known field theory relation between
the energy and the angular momentum. We reach a similar conclusion
for the Klebanov-Strassler background, by showing that the
conical-like configurations are also unstable.

\vfill{
 \hrule width 5.cm
\vskip 2.mm
{\small
\noindent E-mail: pons@ecm.ub.es, pere.talavera@upc.es
}}

\end{titlepage}

\newpage

\section{Motivation}
\setcounter{equation}{0}

The AdS/CFT correspondence constitutes a framework where string
theory and gauge field theories have married successfully. The
conjecture \cite{conjecture1}-\cite{conjecture3} involves the
equivalence between string theory on the bulk of a curved
background, AdS$_5 \times S^5$, and a gauge theory defined on the
boundary, \CN=4 $\SU(N)$ super Yang-Mills theory (SYM). The
duality has been tested in the supergravity approximation, where
the curvature ${\cal R}$ is small and $\alpha^\prime {\cal R}$ can
be neglected. However, any direct relation between string theory
and gauge field theory is extremely difficult to obtain. In fact,
achieving the weak coupling regime in the string sigma-model
becomes a strong coupling regime in the field theory side, where
nowadays our knowledge is restricted to lattice calculations.

One way to overcome the restriction due to the supergravity
approximation, is to find a suitable limit \cite{penrose} where
string theory becomes fully solvable  \cite{ppwave} on NS or RR
backgrounds \cite{metsaev}. Once this limit is achieved in the
string theory side one can indeed apply the AdS/CFT arguments and
look for the corresponding dual field theory states. This
comparison is a overwhelming task, that in the field theory side
is equivalent to solving large-$N$ SYM. It has become
clear only recently that for certain classes of operators
the comparison is still
possible without summing all the $1/N$ series. These correspond to
states having large quantum numbers such as R-charge or spin
\cite{bmn}. In addition, in string theory these states are
stationary and semi-classical \cite{gkp}. Moreover, they do not
encode directly any reference to the strength of the interaction
and therefore they can be used all the
way up to the perturbative regime in the field theory side. In
particular, for large rotating strings in AdS$_5$ one finds the
dispersion relation
\begin{equation}
\label{E-S-QCD}
E-S = \frac{\sqrt{\lambda_{{\rm  AdS}_5}}}{\pi} \ln\left(\frac{S}
{\sqrt{\lambda_{{\rm AdS}_5}}}\right)\,,
\quad \lambda_{{\rm AdS}_5}=g_{\rm YM}^2 N\,,
\end{equation}
for the energy and the spin. Due to conformal invariance $E=\Delta$, with $\Delta$ being the anomalous dimension on $\mathbb{R}^4$. The resemblance with the perturbative result for the
anomalous dimension of twist-two operators in the
conformal gauge theory is striking \cite{qcd}
\begin{equation}
\label{gaugeQCD}
\Delta-S = f(\lambda) \ln S\,.
\end{equation}

The same semi-classical analysis as presented in \cite{gkp} has
been applied to more general cases such as non-supersymmetric or
non-conformal theories \cite{nonconf1}-\cite{nonconf7} in the hope
to gain some understanding on these theories. However, it
is not clear how to properly generalise \bref{E-S-QCD} because,
due to mixing, there is no one-to-one map between states and
operators in non-conformal theories. There has been recent
proposals suggesting as solutions of \bref{gaugeQCD}
non-periodic  solitons
in time-dependent backgrounds
\cite{tseytlin}. It is our aim to generalise in the same spirit as
in \cite{gkp} one of the most relevant \CN=1 SYM model \cite{mn}.
We shall bear in mind \bref{E-S-QCD} and search for a similar
relation. It has been checked that, in the superconformal case
\CN=4, \bref{E-S-QCD} is protected under renormalisation, in the
sense that quantum corrections to \bref{E-S-QCD} proportional to
$\ln^2(\frac{S} {\sqrt{\lambda}})$ cancel out, (see \cite{frolov}
for the string calculation) at least at one-loop level (see also
\cite{n4} for a field theory analogue with R-charge instead of
spin).

We shall explore different settings for classical closed string
configurations, stable in the background of \cite{mn}, and study
relationships between the conserved quantities that arise, like
energy, spin or other quantum numbers. There are still many open
points that deserve clarification, the main one being the exact
string/gauge field duality realisation. Much work must be done
before the fogginess of the present landscape fully dissipates.
Nevertheless, and despite the fact that the dictionary is not yet
ready, we think it is worth to explore one side of the
correspondence --the stringy side-- with the view that our results
must bear some hints on the behaviour of the dual field theory,
yet to be determined.

 The paper is organised as follows. In section \ref{n=1} we
briefly review the \CN=1 background of interest, stressing its
main properties and mainly the role of the Kaluza-Klein (KK)
states. We shall closely follow \cite{mn} in the slightly adapted
notation of \cite{paolo}. In the next sections we explore all the
relevant closed strings configurations: In section \ref{s2} we
consider strings spinning in the $S^2$ and stretching along the
transversal direction, whereas in section \ref{osci} we study
oscillating configurations along the equator of the $S^2$. In
section \ref{secs3} the strings are spinning in the transversal
$S^3$. Conical configurations stretching along a radial variable
in $\mathbb{R}^3$ and also along the transversal direction are
considered in section \ref{secr4}. We devote section \ref{summ} to
summarising our results. Comments on the application of the
variational principle for the Nambu-Goto action describing folded
strings are gathered in the appendix \ref{app}.


\section{\CN=1 Background}
\label{n=1}


We shall consider mainly the semi-classical analysis in the background
produced by a stack of $N$ D5-branes located at the origin of the
transversal coordinate $\rho$, \cite{volkov}, and partially
wrapping a supersymmetric cycle inside a Calabi-Yau three-fold.
The unwrapped sector of the world-volume contains the field theory
in four dimensions, in the infrared \CN=1 SYM. A certain amount of
supersymmetry, namely four supercharges, is kept after the partial
twist. It is also assumed that in this procedure some world-volume
fields become massive enough to decouple.

In the string frame the relevant fields
are given by the metric \footnote{For sake of clarity, we shall work in string units and only restore factors eventually.}
\begin{equation}
\label{metric} ds^2_{10}=e^{\Phi_D} \left[
dx^2_{0,3}+N\alpha^\prime g_s \left(d\rho^2+e^{2g(\rho)} d\Omega_2^2
+\frac{1}{4} (\omega^a-A^a)^2\right) \right]\,,
\end{equation}
the dilaton field
\begin{equation}
\label{dil-D5}
e^{2\Phi_D}=\frac{\sinh2\rho}{2e^{g(\rho)}}\,,
\end{equation}
and the Ramond-Ramond three-form \be F_{[3]}=dC_{[2]}=\frac{N
g_s}{4}\left[-(w^1-A^1)\wedge(w^2-A^2)\wedge(w^3-A^3) +
\sum_{a=1}^3 F^a\wedge(w^a-A^a)\right]\,. \ee
We have introduced
the $\SU(2)_L$ gauge field
\begin{equation}
\label{last2}
 A=\frac{1}{2}\left(\sigma^1 a(\rho) d\theta+\sigma^2 a(\rho)\sin\theta d\varphi
+\sigma^3 \cos\theta d\varphi \right)\,,
\end{equation}
together with its field strength $F^a$, and the left-invariant one-form parameterising
the tree-sphere
\begin{eqnarray}
\omega^1& +& i \omega^2 = e^{-i \psi} \left(d\tilde{\theta}+i\sin\tilde{\theta}
d\phi\right)\,,\nonumber\\
&\omega^2&= d\psi+\cos\tilde\theta d\phi\,.
\end{eqnarray}
In addition
we have also made use of two functions of the radial variable $\rho$, which are given by
\begin{equation}
e^{2g(\rho)}=\rho \coth 2\rho - \frac{\rho^2}{\sinh^2 2\rho}
-\frac{1}{4}\,, \quad a(\rho)=\frac{2\rho}{\sinh 2\rho}\,.
\end{equation}

A good description of the four dimensional field theory is
obtained from the D5-brane in a regime of energies where higher
stringy modes as well as the Kaluza-Klein states on the $S^2$
decouple. It is expected that the back-reaction of the D-brane
deforms the initial Calabi-Yau space \cite{vafa} essentially by
shrinking the $S^2$ and blowing-up the $S^3$. It is then mandatory
to trace back the role played by the point-like Kaluza-Klein modes
in \bref{metric}. The validity of the supergravity approximation
to string theory relies on three conditions:

i) The smallness of the scalar field \bref{dil-D5}. This restricts
the reliability of \bref{metric} to the infrared region.
Surprisingly enough, if one pushes the theory beyond this regime
and calculates the $\beta$-function, one obtains agreement with
the perturbative expression and can even predict non-perturbative
contributions due to fractional instantons
\cite{paolo,Apreda:2001qb}.

ii) The smallness of the curvature. This implies that its maximum
value (attained at the origin) must be bounded
\begin{equation}
\label{curva} \vert {\cal R} \vert \le \frac{32}{3} \frac{1}{N g_s
} \ll 1\,.
\end{equation}

iii) The effective four-dimensional field theory should not be sensible to the
massive Kaluza-Klein modes and therefore their masses should be
heavy enough in order to decouple. This implies the condition
\begin{equation}
\label{kksugra}
N g_s  \ll 1\,.
\end{equation}

Conditions \bref{curva},\bref{kksugra} are incompatible and one is
forced to give up one of them. In section \ref{s2} we shall see
how this claim can be modified once we consider solitonic
string configurations in the supergravity background
\footnote{It is worth
noting that the non-commutative version of this model \cite{mn-nc}
has a region of parameters where the decoupling of the massive KK
modes is compatible with the small curvature condition.}.

The main expectations for the dual field theory, as predicted from the supergravity side,
\bref{metric}--\bref{last2}, are confinement and the correct
$\beta$-function while the most relevant failure is the contamination with
Kaluza-Klein modes which can not decouple in the deep infrared.

We can
also S-dualise the gravity solution and switch to a NS5-brane
description. The field theory side is now replaced by a little
string theory. The metric S-dual to that in \bref{metric} is
\cite{mn}
\begin{equation}
\label{S-metric}
ds^2_{10}= dx^2_{0,3}+N \alpha^\prime g_s \left(d\rho^2+e^{2g(\rho)} d\Omega_2^2
+\frac{1}{4} (\omega^a-A^a)^2\right)\,,
\end{equation}
where the dilaton behaves now as
\begin{equation}
e^{2\Phi}=\frac{2e^{g(\rho)}}{\sinh2\rho}\,,
\end{equation}
and the field strength  $F_{[3]}$ is unchanged.


\section{String rotating in $S^2 \times \mathbb{R}$}
\label{s2}
\setcounter{equation}{0}

We shall consider in this section a spinning string in $S^2 \times
\mathbb{R}$, where $\mathbb{R}$ stands for the $\rho$ variable.
{F}rom the point of view of the field content this corresponds to
the Kaluza-Klein modes associated with the $S^2$, namely the
stringy Kaluza-Klein states in contradistinction to the
supergravity Kaluza-Klein modes (or point-like configurations, see
section \ref{n=1}). In performing a semi-classical analysis the
hope is to gain some insight in their effects and substantiate
their possible decoupling beyond the supergravity approximation.

We shall place a string with fixed coordinates in the equator
of the $S^3$ and in the flat space-time, which is isotropic.
With the remaining variables we consider the classical string
configurations
\begin{equation}
\label{conf1}
t= e \tau\,,\quad \varphi= e \omega \tau \,,\quad \theta(\sigma)\,,\quad
\rho(\sigma)\,,
\end{equation}
with the string rotating around its centre of mass located at
$\rho =0$. Notice that due to the twisting the $\SO(3)$ isometry
in \bref{metric} is broken down to U(1), thus allowing to
interpret the  global charge, $J$, defined by the rotating
configuration \bref{conf1} as the one corresponding to the
R-symmetry of \CN=1 SYM. More specifically we are interested in
the relation $E=E(J)$, where $E$ is just the system energy.
Obviously, instead of \bref{conf1} one can also impose
interpolating configurations involving rotation on both the $S^2$
and the spatial directions of the unwrapped part of the
worldvolume, $\mathbb{R}^3$ \cite{russo}. Besides the R-symmetry,
this will involve the spin. But this will have little to add to
the present discussion on the Kaluza-Klein modes.

In the Nambu-Goto action we shall choose a gauge where $e=1$. The
simplest configuration, with $ \theta$ constant, is only stable
for $\theta=\frac{\pi}{2}$, and the remaining gauge freedom is
absorbed in the variable $\rho$. The Lagrangian is thus given by
\begin{equation}
\label{lag1}
L = -4 \frac{\sqrt{\lambda }}{2\pi} \int_0^{\rho_0} d\rho e^{\Phi_D}
\sqrt{ 1- \lambda \beta(\rho)
\left(\dot{\varphi}\right)^2 }\,, \quad \lambda= N g_s\,,
\end{equation}
where
\begin{equation}
\label{beta}
\beta(\rho) \equiv e^{2g(\rho)} +\frac{1}{4} a^2(\rho)\,,
\end{equation}
and $\dot{\varphi}\equiv d\varphi/d\tau$. The factor of 4 in
\bref{lag1} arises because of the folding of the string. The
string turning point, $\rho_0$, can be found by applying the
variational principle to the Lagrangian \bref{lag1} (see appendix
\bref{app}), and gives the extreme solution for the positivity in
the square root
\begin{equation}
\label{rho11}
1-\lambda \omega^2  \beta(\rho_0) = 0\,,
\end{equation}
which in turn will guarantee a real energy and R-charge. Notice in
particular that at $\rho= \rho_0$, the turning point of the folded
string, the derivative of $\rho$ with respect to $\sigma$ should
vanish. The energy and the R-charge are directly derived from
\bref{lag1},
\begin{eqnarray}
\label{en-spin-NG}
E &=& 4 \frac{\sqrt{\lambda}}{2\pi} \int_0^{\rho_0} d\rho \frac{e^{\Phi_D}}
{\sqrt{ 1- \lambda\omega^2 \beta(\rho)}}\,,\nonumber \\
\omega J &=& 4 \frac{\sqrt{\lambda}}{2\pi} \int_0^{\rho_0} d\rho \frac{\lambda\omega^2
\beta(\rho) e^{\Phi_D}}
{\sqrt{ 1- \lambda\omega^2 \beta(\rho)}}\,.
\end{eqnarray}

It is worth re-discussing the above expressions in the conformal
gauge. First of all because it clarifies whether a particular
ansatz fulfils all the requirements imposed by the equations of
motion. This can certainly be traced back in the Nambu-Goto action
but is somewhat cumbersome. Secondly because it will be a good
check of the restriction imposed in the integration range of the
$\rho$ variable. The world-sheet action is in this case
\begin{equation}
\label{lag2}
S=-\frac{1}{4\pi} \int d\tau d\sigma G_{ij} \partial_\alpha X^i \partial^\alpha X^j\,,
\end{equation}
that must be supplemented with the conditions

\begin{equation}
\label{conss} T_{\alpha\beta} = \partial_\alpha X^i
\partial_\beta X^j G_{ij} - \frac{1}{2} \eta_{\alpha\beta}
(\eta^{\gamma\delta}\partial_\gamma X^i
\partial_\delta X^j G_{ij})=0\,.
\end{equation}
Notice that the components of the energy-momentum tensor
$T_{\alpha\beta}$ are constants of motion for the conformal gauge
action \bref{lag2} \footnote{They must vanish because they were
constraints for the Polyakov action before implementing any gauge
fixing (the Lagrangian in the conformal gauge \bref{lag2} is
regular and therefore has no constraints of its own).}. It is also
worth mentioning that the choice $\theta = \pi/2$ we picked out in
the Nambu-Goto action simply arises in the conformal gauge from
the equation of motion for $\theta$. For the string configuration depicted in
\bref{conf1} and $\theta = \pi/2$, the contents of these
constraints becomes the single relationship
\begin{equation}
\label{constrain1}
\lambda (\rho^\prime)^2 = e^2 \left(1 - \lambda \omega^2 \beta(\rho) \right)\,,
\end{equation}
that reduces to \bref{rho11} once we set $\rho^\prime\vert_{\rho=\rho_0}=0$.
Notice that this identifies the turning point obtained above within the
Nambu-Goto approach.
Hereafter primes stand for derivatives with respect to $\sigma$.
The constant $e$ is adjusted in order to get a period of $2\pi$ in the
function $\rho(\sigma)$
\begin{equation}
d\sigma = d\rho \,\,\frac{\sqrt{\lambda}}{e \sqrt{1-\lambda\omega^2 \beta(\rho)}}\,.
\end{equation}
Using \bref{lag2} and \bref{conf1} one obtains the space-time energy and spin
in terms of $\sigma$
\begin{eqnarray}
\label{en-spin-CG}
E &=&  \frac{1}{2\pi} e \int d\sigma\, e^{\Phi_D} \,,\nonumber \\
\omega J &=&  \frac{\lambda}{2\pi} e \omega^2 \int d\sigma\,
e^{\Phi_D} \beta(\rho)\,,
\end{eqnarray}
that with the use of \bref{constrain1} reduce to the previous
expressions \bref{en-spin-NG} \footnote{Since the parameter $e$ is
usually different for unity in the conformal gauge, to get an
interpretation of \bref{en-spin-CG} as space-time quantities, one
must take the Lagrangian such that the action is expressed as
$S=\int dt L$, with $t$ instead of $\tau$.}.

To get some elementary information we analyse in particular the
limiting cases of long and short strings.

 {\bf Long strings:} We shall first discuss long closed strings \cite{gkp}.
We expect that the string feels the $S^2$ curvature and hence a
change in the $E(J)$ relation with respect to the flat case would
be induced. In the ultraviolet \bref{beta} can be approximated by
$ \beta(\rho)\approx \rho \,, $ and the dilaton term by
$e^{\Phi_D} \approx \frac{1}{2} \rho^{- \frac{1}{4}} e^\rho$. This
would suffice in the present situation to display non-analytical
terms. The situation corresponds, from \bref{constrain1}, to
approach $\omega\rightarrow0^+$, in contrast to the AdS$_5$ case,
$\omega\rightarrow1^-$. Hence $\rho_0$ becomes large. Within this
approximation \bref{en-spin-CG} can be expressed in terms of
hypergeometric functions and reduces to
\begin{eqnarray}
\label{E-S-UV-D5branes} E-\omega J&=&
\frac{\sqrt{\lambda}}{\pi}\frac{\sqrt{\pi}}{2}
\frac{\Gamma(\frac{3}{4})}{\Gamma(\frac{9}{4})} \rho_0^{3/4}
~_1F_1(\frac{3}{4},\frac{9}{4};\rho_0) \underset{\rho_0
\gg}{\rightarrow} \frac{1}{2} \sqrt{\frac{\lambda}{\pi}}
\rho_0^{-3/4} e^{\rho_0}\,,\nonumber \\
\omega J &=& \frac{\sqrt{\lambda}}{\pi}\sqrt{\pi}
\frac{\Gamma(\frac{7}{4})}{\Gamma(\frac{9}{4})} \rho_0^{3/4}
~_1F_1(\frac{7}{4},\frac{9}{4};\rho_0) \underset{\rho_0
\gg}{\rightarrow} \sqrt{\frac{\lambda}{\pi}} \rho_0^{1/4}
e^{\rho_0}\,,
\end{eqnarray}
where in the last step we have just kept the leading contribution
in the $~_1F_1(a,b;\rho_0)$ function as $\rho_0 \rightarrow
\infty$. Both operators, $E$ and $J$ diverge with $\rho$ while,
similar to the AdS$_5$ case, the ratio $\frac{E-\omega J}{\omega
J}$ remains finite. In this approximation we may use, in string
units, $$ \omega^2 = \frac{1}{\lambda\rho_0} =
\frac{1}{R^2\rho_0}\,, \quad R^2 = \lambda \alpha^\prime = N
\alpha^\prime g_s \,,
$$
and we have the leading terms for $E$,\ $\omega J$ and $E-\omega
J$,
$$ E \approx \omega J =
\frac{R}{\sqrt{\pi}}\rho_0^{1/4}e^{\rho_0}\,,\quad
E-\omega J =
\frac{1}{2}\frac{R}{\sqrt{\pi}}\rho_0^{-3/4}e^{\rho_0}\,.
$$
Keeping the next-to-leading order in $\rho_0$
and since $\rho_0 \approx \log J$ for large $\rho_0$, we can
write this expression up to the subleading term as
\begin{equation}
\label{eJ} E=\frac{J}{R \log^{1/2} J} + \frac{1}{2} \frac{J}{R
\log^{3/2} J}+\dots\,.
\end{equation}
where the ellipsis stands for higher terms in $R$. The dependence
on the R-charge $J$ in this last expression shows that, keeping
ourselves in the supergravity approximation, as long as $J$ is
sufficiently large whilst $R$ is keep fixed, the energy of the KK
stringy modes increases and their effects decouple. The content of
\bref{eJ}, found in the string theory side, has no direct
reference to $g_s$ (only to the product $N g_s$) and its validity
is beyond the weak coupling regime. Furthermore it still holds at
strong coupling and therefore, even if to the best of our
knowledge it is not known, it should have an analogous expression
in the \emph{perturbative} gauge field context, as \bref{E-S-QCD}
has its equivalent \bref{gaugeQCD} \cite{gkp}.

It is distressing that a rescaling of $R$ yielding (for fixed
$g_s$) $N\rightarrow \infty$ can change drastically the
aforementioned pattern concerning the KK decoupling. The
$R\rightarrow \infty$ limit can be achieved, for instance, by
transforming the \bref{metric} geometry into a parallel plane-wave
space-time \cite{gimon}.

Let us inspect which are the consequences as far as the KK decoupling
is concerning by going to the parameter corner
\begin{equation}
\label{limit}
R\rightarrow \infty\,,J\rightarrow \infty\,,\quad
\end{equation}
with $\frac{J}{R}$ fixed but otherwise free, either large or
small.
By virtue of \bref{limit}, and bearing in mind that we have just
argued that \bref{eJ} holds at any $g_s$, this implies
$N\rightarrow \infty$. The value of $\lambda_{\rm t'Hooft} =
g^2_{\rm YM} N$ can be read directly from the behaviour of the
$\beta$-function \cite{olesen,bertolini}
\begin{eqnarray}
\frac{1}{\lambda_{\rm t'Hooft}} &\sim& \frac{g_s}{4 \pi^2}\rho^2
\quad {\rm
for}\quad\rho\rightarrow 0\,,\nonumber\\
\label{betapaolo}
\frac{1}{\lambda_{\rm t'Hooft}} &\sim& \frac{g_s}{4
\pi^2}\rho\quad {\rm for}\quad\rho\rightarrow \infty\,,
\end{eqnarray}
diverging in the infrared and vanishing in the ultraviolet. In the last case
$g_{\rm YM}^2$ tends to zero sufficiently fast, whilst $N$ is taken large,
ensuring the
applicability of perturbation theory in the field theory dual.
While the former implies a slow decreasing of $g_{\rm YM}^2$ when approaching
the infra-red region and the field theory dual becomes strongly coupled.
Thus, after the double scaling \bref{limit} we have neither recovered
after all the t'Hooft limit, i.e. $N \rightarrow \infty\,,
g^2_{\rm YM}\rightarrow 0$ with $\lambda_{\rm t'Hooft} = g^2_{\rm
YM} N$ fixed
nor the superconformal case
equivalent of \bref{limit}, BMN limit, where \cite{bmn}
\begin{equation}
\label{limitbmn} R\rightarrow \infty\,,J\rightarrow \infty\,,\quad
{\rm with}\quad \frac{J^2}{R_{AdS_5\times S^5}} \quad {\rm and}
\quad g^2_{\rm YM} \quad {\rm fixed}
\end{equation}
with the only restriction $g_{\rm YM}^2 \ll 1$ but otherwise not
vanishing \footnote{Contrary to \bref{limit}, in \bref{limitbmn}
$g_{\rm YM}$ can remain fixed, for a given $g_s$, because in the
superconformal case $g_{\rm YM}^2 = 4 \pi g_s$ without involving
any factor of $N$.}. It is evident that this last large-N limit is
different from the usual t'Hooft limit. One can argue that even if
we have previously found a decoupling of the KK modes, it is not
so obvious that this holds true in all the parameter space
$(R,J)$, see the scaling \bref{limit}. However, one can not rule
out a priori that a scaling of the type $J> R$ will not still make
plausible a decoupling of the Kaluza-Klein modes even in this
corner, $R\rightarrow \infty$.

One can argue that the description of the D5 brane background in
the ultraviolet region is not suitable for physical purposes
because the blowing up of the dilaton. To elucidate further the
decoupling of the Kaluza-Klein states we repeat the calculations
in the S-dualised metric \bref{S-metric}. Since our configuration
does not couple to the NS-NS background field, the action to
consider is again Nambu-Goto. It is convenient to switch to the
following set of variables \cite{Kruczenski:2002fb}
\begin{eqnarray}
\label{pp-pm}
P_+=E + \omega J &=& 4 \frac{\sqrt{\lambda }}{2\pi} \int_0^{\rho_0} d\rho
\frac{ 1+ \lambda \omega^2
\beta(\rho)}
{\sqrt{ 1- \lambda \omega^2 \beta(\rho)}}\,, \nonumber\\
P_-=E -\omega J&=& 4 \frac{\sqrt{\lambda }}{2\pi} \int_0^{\rho_0} d\rho
\sqrt{ 1-  \lambda \omega^2 \beta(\rho)}\,,
\end{eqnarray}
that would allow a non-trivial combination of the operators $E$
and $J$. In the present case, and because the function
$\Phi_D$ is not present in the metric, we should retain the
full expression for $\beta(\rho)$ in obtaining the turning point,
which for large values of $\rho$ is given by
\begin{equation}
 \lambda \omega^2 = \frac{1}{\rho_0 \coth(2\rho_0)}\,.
\end{equation}
Expanding \bref{pp-pm} around $\rho_0$, and subtracting the
appropriate combination on $P_+, P_-$ in order to cancel the
leading term one finds
\begin{equation}
P_+-5P_- = 4 \frac{\sqrt{\lambda }}{2\pi} \frac{\rho_0^2 {\rm
csh}(\rho_0){\rm sech}(\rho_0)} {\sqrt{1-\rho_0 {\rm csh}(\rho_0)
{\rm sech}(\rho_0)}} \underset{\rho_0 \gg}{\rightarrow} 8
\frac{\sqrt{\lambda }}{\pi} \rho_0^2 e^{-2\rho_0}\,.
\end{equation}
Using the leading contribution to $P_+$ or $P_-$
\begin{equation}
P_+ = 5 P_- \approx 4 \frac{\sqrt{ \lambda }}{2\pi} \frac{10}{3}\rho_0\,,
\end{equation}
we arrive finally at the closed expression
\begin{equation}
\label{eJs}
E=\left(\frac{3}{\sqrt{\pi}}\right)^{2/3}\left(\frac{J}{\sqrt{R}}\right)^{2/3}+\ldots\,.
\end{equation}
We see again from this expression, similarly to \bref{eJ},
that the energy increases with $J$, signalling the decoupling of
the KK stringy states with sufficiently large values of the
R-charge.

In addition we can inspect the behaviour of the KK states once the
parallel plane-wave limit for \bref{S-metric} is taken. As
previously the condition $R\rightarrow \infty$ arises. A similar
argument as has been presented previously indicates that the same
kind of scaling, $J \sim \sqrt{R}$ still holds here and would
probably suffice for the decoupling.

Summarising, we can conclude that, contrary to the Kaluza-Klein
point-like modes, the KK stringy modes on the $S^2$ decouple for
states with sufficient large quantum-number $J$. This must be the
case from the very beginning not only for the stringy KK modes but
also for higher realisations: as the field theory dual does not
posses an explicit U(1) symmetry the states corresponding to this
charge should decouple. This means that although there is the
presence of KK modes that prevents the supergravity model from
being the exact dual of ${\cal N} =1$ SYM for large $N$, the
extent of this KK contamination is lighter than it could have been
expected in view of the incompatibility between \bref{curva} and
\bref{kksugra}. However, the relation for $E(J)$ that we find for
the $S^2$ stringy modes is not of the \bref{E-S-QCD} type, neither
we think it is forced to be, because it has nothing special to do
the field theory side in the noncompact dimensions. Surprisingly
enough this is not the case for other backgrounds
\cite{Hartnoll:2002th} where from a similar treatment it seems
possible to obtain \bref{E-S-QCD}, although is not clear to us its
interpretation. If in addition we also perform the parallel
plane-wave limit, we are forced to consider a double scaling limit
($R,J\rightarrow \infty$) that neither ends in the classical
t'Hooft limit nor in the BMN one. Under certain circumstances this
limit can lead also to a possible decoupling. However, in this
last case, it is less clear to us the role played by non-planar
diagrams in the dual field theory side. Probably and by analogy
with the AdS case, the choice $J\sim R$ for the D5-brane system is
the critical situation where non-planar diagrams in the field
theory side are neither dominant nor subdominant with respect to
the planar ones \cite{charlotte}.

\hspace{1cm}

{\bf Short strings:} The situation for small $\rho_0$ values
implies large $\omega$. We expect a similar behaviour to that in
flat Minkowski space-time. Furthermore the role of the dilaton is
higher order in $\rho$ and thus both, D5-brane and NS5-brane,
backgrounds should give the same results. As before we take a
spinning string with the configuration \bref{conf1}. Expanding
\bref{pp-pm} around $\rho_0 \approx 0$ we get the leading
behaviours
\begin{equation}
E =  \omega J = \sqrt{\frac{3}{2}} \frac{1}{\omega}\,,
\end{equation}
so that
\begin{equation}
\label{e2-short}
E^2 = \sqrt{\frac{3}{2}}  J
\end{equation}
Expression \bref{e2-short} should be compared with the flat case
 $E^2 = 2 J$.
Between both relations there is a mismatch of a factor
$\sqrt{\frac{8}{3}}$, thus we are not getting exactly flat space.
This factor is easily understood from the infra-red asymptotics of
\bref{metric} [cf. \bref{S-metric}]. At small $\rho$,
$\beta(\rho)\rightarrow \frac{1}{4}$, whereas the analogous
expression in flat space tends to zero. To be more definite, even
if the prefactor $e^{2 g(\rho)}$ of the $S^2$ vanishes in this
limit, the contribution of the $a(\rho)$ field remains --due to
the twisting-- preventing the $S^2$ from eventually shrinking.
Probably this can be amended by using the $S^2$ parametrisation of
\cite{bertolini}.

\section{String oscillating in $S^2 \times \mathbb{R}$}
\label{osci} \setcounter{equation}{0}

It has become more or less clear that all the metrics which
possess a conformal invariance behave similarly when oscillating
strings are considered. The model in \cite{mn} lacks of the
presence of this symmetry and hence to get a more elementary
information on the role played by conformal invariance, 
or its absence thereof, we consider a multiwrapped closed string
which oscillates around the centre of the $S^2 \times \mathbb{R}$.

The motion of the string we are describing is a
pulsation that reaches a maximum for $\rho$ at some $\rho_{\max}$,
then the string shrinks, collapses for $\rho =0$ and expands again
until $\rho_{\max}$. The explicit configuration of this
\emph{pulsating} string is
\begin{equation}
t = \tau\,,\quad \varphi(\sigma)\,,\quad \rho(\tau)\,,\quad \
\theta = \frac{\pi}{2}\,, \label{configuration}
\end{equation}
where $\varphi(\sigma + 2\pi)= \varphi(\sigma) + 2\pi m,$ denoting
$m$ the number of times the string wraps the $\varphi$ direction.
Note that we could in principle have taken a seemingly more
general configuration, $\varphi(\tau,\sigma)$, but the geometrical
meaning of the Nambu-Goto action for the string makes it
indistinguishable from the one derived from \bref{configuration}.

The Nambu-Goto action restricted to the configuration
(\ref{configuration}) is
\begin{eqnarray}
\label{oscc} S
&=& -m \int d\tau  e^{\Phi_D} \sqrt{ (1- \lambda \dot{\rho}^2)
\lambda \beta(\rho)} \,,
\end{eqnarray}
which defines a one-dimensional classical system to which we can
apply the usual canonical analysis. This immediately leads to the
momenta
\begin{equation}
\label{momenta} \Pi_\rho= e^{\Phi_D} \frac{\lambda^2 m\dot{\rho}
\beta(\rho)} { \sqrt{(1- \lambda \dot{\rho}^2)\lambda
\beta(\rho)}}\,,
\end{equation}
and the -squared- Hamiltonian
\begin{equation}
\label{hamil} {\cal H}^2 = e^{2 \Phi_D} \frac{\lambda m^2
\beta(\rho)}{1- \lambda \dot{\rho}^2}\,.
\end{equation}
{From} \bref{hamil} and \bref{momenta} one obtains
\begin{equation}
\label{hnew}
{\cal H} = \sqrt{ \frac{\Pi_\rho^2}{\lambda}+ e^{2 \Phi_D} \lambda
m^2 \beta(\rho)}\,,
\end{equation}
which defines a one-dimensional potential
\begin{equation}
\label{pot} V(\rho) =  e^{2 \Phi_D} \lambda m^2 \beta(\rho)\,.
\end{equation}

In what follows we shall find the energy eigenvalues using the WKB
approximation and compare with \cite{nonconf6}. These authors
suggest for non-conformal AdS backgrounds in the high-energy limit
a \emph{universal} behaviour
\begin{equation}
\label{bsnaive} E\sim n + f(\lambda) \sqrt{m n }\,,
\end{equation}
where $n$ is the principal quantum number and $f$ refers to any
function of the coupling constant.

The WKB expression gives
\begin{equation}
I \equiv \int_0^{\rho_{\rm max}} d\rho \sqrt{\lambda E^2 - e^{2
\Phi_D} \lambda^2 m^2 \beta(\rho)}= \left(n+\frac{1}{2}\right)\pi
\,.
\end{equation}
Notice that the potential \bref{pot} grows exponentially as $\rho$
increases, thus $E$ large implies inherently $\rho$
large. Under this approximation
\begin{equation}
\label{sine} I\approx \frac{\sqrt{\lambda}}{2} E
\ln\left(\frac{4E^2}{\lambda m^2}\right) + {\cal
O}\left(\frac{1}{E}\right)\,,
\end{equation}
expression that matches with the energy levels (with $\sqrt{E}$
instead of $E$) of a N=2 super Sine-Gordon model
\cite{russo-tseytlin}. At this point we can not tell whether this matching
is coincidental or relies on more fundamental grounds. If the
latter were the case, our results could even suggest that in the
deep ultraviolet \bref{oscc} defines an integrable system. By
passing notice that \bref{sine} is not like as \bref{bsnaive},
therefore \bref{oscc} does not belong to the same equivalence
class of the AdS models at finite temperature.

In addition we have explored a second limit, the deep IR, where we
demand $\sqrt{\lambda} E$ to be small. Then the string is
restricted to be close to the origin of the radial variable $\rho$
and therefore, as in the short string case we discussed
previously, in principle we expect it not to feel the curvature.
The potential near the origin can be obtained by expanding
\bref{hnew} as a power series and looks like that of a shifted
harmonic oscillator
\begin{equation}
V(\rho) = \lambda^2 m^2 \left(1+\frac{20}{9}\rho^2\right)\,.
\end{equation}
Then the squared-energy turns out to be proportional to the
excitation level of the oscillator, given by the product $m n$,
\begin{equation}
E^2_n - E^2_0 \approx \sqrt{\lambda} m n\,,
\end{equation}
that, as expected, matches with the spectrum of the pulsating string in flat space
$
E^2_n - E^2_0 \approx 4 m n\,.
$


\section{String rotating in $S^3$}
\label{secs3}
\setcounter{equation}{0}

Instead of decoupling the $S^3$ by just freezing its coordinates
as has been done in the previous section one can similarly place a
spinning string with the centre of mass located at the north pole
of the $S^3$ and at the origin of the transversal
coordinate \footnote{The configuration is only stable at
$\rho=0$ because this is the only point for which $\frac{d
\phi_D}{d \rho}=0$.} and study its states. These should
correspond to highly excited string states \cite{gkp}. It  should
also reinforce the idea that the holographic coordinate $\rho$ is
an essential ingredient to obtain a non-trivial behaviour in the
$E(J)$ relation. Let us write the $S^3$ metric coming from
\bref{metric} as
\begin{equation}
\label{metric3}
d\Omega_3^2 = \frac{1}{4}\lambda \sum_a
\left(w^a\right)^2=\frac{1}{4} \lambda \left( d\tilde{\theta}^2
+\sin^2\tilde{\theta} d\tilde{\phi}^2 +\cos\tilde{\theta}^2
d\tilde{\varphi}^2\right)\,,
\end{equation}
where we have changed to a different set of coordinates for the
$S^3$ (see the appendix in \cite{paolo}). The notation for the
angular coordinates is exclusive for this section. We shall adopt
the ansatz
\begin{equation}
\label{conf3}
t=e \tau\,,\quad \tilde{\theta}(\sigma)\,,\quad
\tilde{\phi}= e \omega \tau\,,\quad \tilde{\varphi}={\rm constant}\,.
\end{equation}
A simple inspection of \bref{metric3} under the restriction
\bref{conf3} reveals the similitude with the case of a spinning
string in the $S^5$ discussed in \cite{gkp} where the AdS$_5$
radius is fixed to $R^2= \frac{1}{4} \lambda$. Therefore we recall
the findings in \cite{gkp} stressing some of their features.

Working out in the conformal gauge it can be checked that
\bref{conf3} is consistent with the equations of motion and the
constraint \bref{conss}. The latter comes to be
\begin{equation}
\label{xx}
R^2 \left(\tilde{\theta}^\prime\right)^2 = e^2 \left( 1 - R^2 \omega^2 \sin^2\tilde{\theta}\right)\,,
\quad R^2= \frac{1}{4} \lambda\,,
\end{equation}
and gives rise to the bound $R \omega \ge 1$.
The exact value of the constant $e$ is obtained by demanding a
period $2\pi$ in the variable $\sigma$. It is rather
straightforward with the help of \bref{xx} to obtain the
expressions for the space-time energy and the $\SO(4)$ quantum
number in terms of the turning point $\tilde{\theta}_0$ (the
string is now stretched along the $\tilde{\theta}$ direction)
\begin{eqnarray}
E&=&\frac{1}{2\pi}e\int_0^{2\pi} d\sigma = 4 \frac{R}{2\pi}
F(\tilde{\theta}_0\vert R^2\omega^2)\,,\nonumber\\
J&=&\frac{R^2}{2\pi}e\omega \int_0^{2\pi} d\sigma
\sin^2\tilde{\theta} = 4 \frac{R}{2\pi \omega} \left\{
F(\tilde{\theta}_0\vert R^2\omega^2)-E(\tilde{\theta}_0\vert
R^2\omega^2) \right\}\,,
\end{eqnarray}
with
\begin{equation}
\sin\tilde{\theta}_0=\frac{1}{R \omega}\,,
\end{equation}
and where we have make used of the elliptic integrals of 1st and
2nd kind, $F$ and $E$ respectively. In the case at hand the regime
of interest, large $J$, is obtained in the limit $R \omega
\rightarrow 1^+\,, \tilde{\theta}_0\sim\pi/2$. Then both
quantities are large  while the
difference $E-\omega J$ remains finite
\begin{equation}
E-\omega J\rightarrow 4 \frac{R^2}{2\pi}=\frac{\lambda}{2\pi}\,.
\end{equation}

\section{String rotating in $\mathbb{R}^3$}
\label{secr4}
\setcounter{equation}{0}

So far we have dealt entirely with the transverse coordinates to
$\mathbb{R}^3$. It is natural to ask whether is possible to
obtain a relation similar to \bref{E-S-QCD} by using the flat
and the holographic coordinates. Naively, one would think that
is quite unlikely, because in the string frame the flat part,
$\mathbb{R}^4$, of \bref{metric} is trivial in the sense that the
transverse coordinate $\rho$ enters as a global common factor to
all the coordinates. In some sense to obtain a non-trivial
relation similar to \bref{E-S-QCD} involves different weighting in
some of these coordinates.

\subsection{Flat $\mathbb{R}^3$}
\label{flatr4}

It is worth, before proceeding with the analysis of \bref{metric},
to clean up the field in the simplest case: plain flat
space-metric. Due to the special way in which the transverse
coordinate $\rho$ comes in \bref{metric} we shall learn in this
manner almost all the needed features.

As we showed above it is equivalent to use the Nambu-Goto action
or the conformal gauge formalism. The analysis will be performed
in the former, but the same results can be attained in the latter.

\begin{figure}[t]
\begin{center}
\epsfig{file=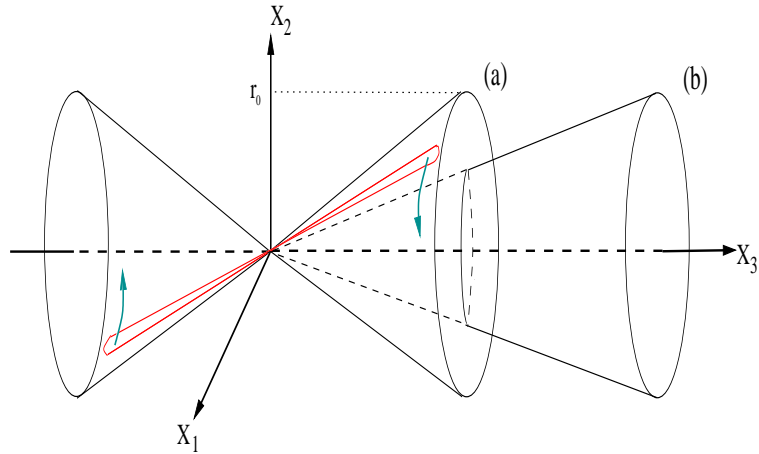,width=10cm,height=6cm}
\end{center}
\caption{Conical configuration for a closed folded string in flat
space. Keeping its centre of mass at the vertex of the cone, the
string stretches, at a given time, along the generator of the
cone, rotating in such a way that its extremes --the points where
the folding takes place at the boundary of the cone-- move at the
speed of light. Configurations a) (complete) and b) (incomplete)
are distinguished because the stretching along the $x_3$ direction
is different. } \label{fig}
\end{figure}

Let us consider a closed string with, for stability reasons, its
centre of mass static and fixed at some arbitrary point that in
the remainder should be taken as the origin of coordinates. We
take cylindrical coordinates $r,\, \varphi,\, x_3$ in
$\mathbb{R}^3$. The string, folded, spins over the surface of a
cone as depicted in fig. \ref{fig} with the configuration
\begin{eqnarray}
\label{flat-conf1}
t = \tau\,, \quad r(\sigma) \,, \quad \varphi = \omega \tau  \,,
\quad x_3(\sigma) \,.
\end{eqnarray}

The Lagrangian density is given by ($L = \int d\sigma {\cal L}$)
\begin{equation}
\label{lagflat}
{\cal L}=-\frac{1}{2\pi} \sqrt{\left(1-r^2 \omega^2\right)\left( (r^\prime)^2
+ (x_3^\prime)^2\right)}\,,
\end{equation}
where we have already substituted $\omega$ for $\dot \varphi$,
which is a trivial solution of the equations of motion for the
time evolution. The equations for the variables $r$ and $x_3$,
depending on $\sigma$, yield two constants of motion,
\begin{eqnarray}
\label{ct}
X[r,x_3] (r^\prime)^2 + r^2 \omega^2 = c_1\,,\quad
X[r,x_3] (x_3^\prime)^2 = c_2\,.
\end{eqnarray}
where
\begin{equation}
X[r,x_3] = \frac{ 1-r^2 \omega^2}{(r^\prime)^2  + (x_3^\prime)^2} \,.
\end{equation}
A similar argument to the one developed in the appendix \bref{app} can be used now
to show that the turning points for the folded string must be located at $r$
satisfying $1-r^2 \omega^2 = 0$. Combining the two constants of motion one gets,
after some manipulations,
\begin{equation}
(1-c_1-c_2) (1-r^2 \omega^2) (r^\prime)^2 = 0 \,,
\end{equation}
and since we expect a configuration where the factors $(1-r^2
\omega^2)$ and $(r^\prime)^2$ can be simultaneously different from
zero, we conclude that $c_1+c_2 = 1$. On the other hand notice
that whereas the freedom of $\tau$ reparameterisations has been
already fixed with $t=\tau$, the freedom of $\sigma$
reparameterisations is still with us. We use it by fixing $X=1$.
Then the second constant of motion reads $(x_3^\prime)^2 = c_2$,
but since the configuration must be periodical in $\sigma$, we end
up with $c_2 =0$ and $x_3= x_{3_0}$, where $x_{3_0}$ is an
arbitrary constant. As for the $r$ variable, we have a standard
oscillatory dependence on $\sigma$, $r(\sigma) =
\sin(\omega\sigma)$. Therefore the conical configuration is not
stable and we end up with a spinning string in flat two
dimensional space.

One can indeed use a more general decomposition in \bref{flat-conf1}, for instance
\begin{eqnarray}
t = \tau\,, \quad r(\sigma)\,, \quad \varphi =\omega \tau\,, \quad x_3(\tau\,,\sigma)
\end{eqnarray}
The result turns out to be identical to the previous one, but with
the plane of motion (coordinates $r, \varphi$) boosted along the
$x_3$ direction. Let us mention also that considering
configurations like $t = \tau + \tilde{t}(\sigma)$ \cite{loewy-oz}
does no changes our conclusions as regards to their
stability.

To summarise, we have learnt that in flat space-time there can not
exist stable string configurations involving the transverse
coordinate, and the strings are confined to rotate in a plane. Is
this behaviour changed by the flatlessness of \bref{metric}\,?
Essentially the main difference between the flat case an the
$\mathbb{R}^3$ part depicted in \bref{metric} is the presence of
the global factor $e^{\Phi_D}$ in the metric. Then it becomes
clear that in the deep infrared this is not the case because the
effect of this factor disappears and we recover flat Minkowski
space-time. We shall turn now to the full analysis of
\bref{metric}.


\subsection{ $\mathbb{R}^3\times \mathbb{R}$ in \CN=1 SYM}

Now $\mathbb{R}$ stands again for the $\rho$ variable. We shall
focus the analysis in \bref{metric}, and contrary to the previous
section we shall employ the conformal gauge formalism, but once
more the Nambu-Goto action leads to the same conclusions, although
it is more messy to use in the equations of motion.

The equations of motion for the --consistent-- configuration
\begin{equation}
\label{conflast} t = e \tau\,,\quad \varphi= e \omega \tau\,,\quad
r(\sigma)\,,\quad \rho(\sigma)\,,
\end{equation}
are
\begin{equation}
\label{eq-r}
\partial_\sigma\left(e^{\Phi_D} r^\prime \right) + e^{\Phi_D} r e^2 \omega^2 = 0\,,
\end{equation}
\begin{equation}
\label{eq-rho}
2 \partial_\sigma\left(N g_s e^{\Phi_D} \rho^\prime \right) - \frac{\partial e^{\Phi_D}}{\partial \rho}
\left( e^2-r^2 e^2 \omega^2+ (r^\prime)^2 + N g_s (\rho^\prime)^2 \right) = 0\,,
\end{equation}
while the rest of the equations of motion are satisfied trivially.
In addition the equation for the constraint leads to
\begin{equation}
\label{cons}
(r^\prime)^2 + N g_s (\rho^\prime)^2 = e^2 \left( 1 - r^2 \omega^2 \right)\,.
\end{equation}
It is quite straightforward to reduce \bref{eq-rho} to \bref{eq-r}
with the help of \bref{cons} and therefore in the remainder we
shall only deal with two equations.

To begin with let us see that $r(\sigma)$ and $\rho(\sigma)$ can not be related via
a continuous
transformation. Suppose $\rho=$ \CH$(r)$ then \bref{eq-r} and \bref{cons} gives
\begin{equation}
\label{eq-r4}
\left(1-r^2 \omega^2\right) \frac{ \dot{{\cal H}}}{1+\dot{{\cal H}}^2}
\frac{d \Phi_D}{d{\cal H}}
-\frac{1}{2} \frac{\partial}{\partial r}\left[\left(1-r^2 \omega^2\right)
\frac{ \dot{{\cal H}}}{1+\dot{{\cal H}}^2} \right] =0\,.
\end{equation}
Although to the best of our knowledge \bref{eq-r4} can not be
solved analytically we have checked numerically that there is no
possible solution if in addition we implement the boundary
conditions $\rho^\prime(r_0) =0, \rho(r_0)={\rm constant}$. The
first one is equivalent to demanding that the $\rho$ edge be at
$r=r_0$. This is a quite plausible condition in view of
\bref{cons}. The statement of non-existence of solution is
independent on the second condition.

The picture we have achieved is even clearer by examining directly
\bref{eq-r} and \bref{eq-rho}, that is, by turning to the most
general case. The equations of motion can be written as
\begin{eqnarray}
\label{eq-r1}
r^{\prime\prime} + r e^2 \omega^2 + \frac{d\Phi_D}{d\rho} r^\prime\rho^\prime=0\,, \\
\label{eq-rho1}
N \rho^{\prime\prime} - \frac{d\Phi_D}{d\rho} (r^\prime)^2 =0\,.
\end{eqnarray}
The variables $r$ and $\rho$ are "radial" in the sense that they
are limited to positive values, but for the true content of these
equations it is best to consider them extended to negative values
as well (in fact, this has been already done without previous
warning in the preceding section for the variable $r$ in the flat
case). Now the only addition that must be made is to take the
function $\Phi_D$ as an even function of the variable $\rho$ which
indeed corresponds to its analytical continuation. It is then easy
to realize that, since $\frac{d\Phi_D}{d\rho} \geq 0$ for $\rho
\geq 0$, equation \bref{eq-rho1} implies that $\rho(\sigma)$ can
never be a periodical function except for the trivial
configuration with $\rho =0$ --that makes
$\frac{d\Phi_D}{d\rho}(0) =0$. In some sense it is as we had the
"wrong sign" for $\Phi_D$. As will be explained below, if the
function $\Phi_D$ in \bref{metric} had been defined with the
opposite sign, we would have found stable conical-like
configurations.


\begin{figure}[h,t]
\begin{center}
\epsfig{file=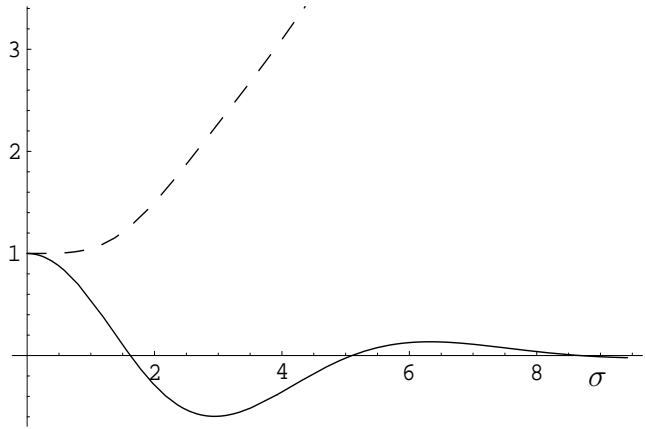,width=9cm,height=6cm}
\end{center}
\vspace{-1.7cm}\hspace{11.5cm}$\sigma$
\vspace{1.7cm}\hspace{-11.5cm} \caption{Behaviour of $r(\sigma)$
(full line) and $\rho(\sigma)$ (dashed curve) for a standard
solution of \bref{eq-r1} and \bref{eq-rho1}. The variable $r$
describes a damped oscillator (the damping is induced by the last
term in \bref{eq-r1}) whereas the variable $\rho$ tends to a
configuration with constant slope $\rho^\prime$, never
periodical.} \label{sol}
\end{figure}

Thus, for $\rho=0$ we are back to the solution of a string
rotating in a plane in flat space. It is worth commenting a little
more on the general solution for the equations \bref{eq-r1} and
\bref{eq-rho1}. The second equation is highly non-trivial once
\bref{dil-D5} is used but in its actual form can be seen to
represent a forced oscillator. The last term in the first equation
\bref{eq-r1} acts as a damping factor with a non-constant
coefficient for the term linear in the velocity $r^\prime$,
$b(\sigma)= \frac{d\Phi_D}{d\rho} \rho^\prime(\sigma)$. Notice
that, with the initial conditions in $\sigma$ described below,
from \bref{eq-rho1} we can conclude that $\rho^\prime(\sigma)$
(and $\rho(\sigma)$) is a monotonically increasing function.
Combining this fact together with the positivity property
$d\Phi_D/d\rho \ge 0$  we can infer that the $b(\sigma)$
coefficient is positively defined, and as a consequence the
amplitude in $r(\sigma)$ decreases as we increase $\sigma$. One
obtains a critical or over-critical oscillator depending on the
exact boundary conditions. Moreover, as we search for periodic
solutions in $r$, $r(\sigma)=r(\sigma+2\pi)$, we only shall
consider the former case. Together with \bref{eq-r1} we have
implemented the following set of initial conditions: at some
$\sigma_0$ corresponding to $r=r_0$, both coordinates, $r$ and
$\rho$ acquire an extremal value, a fact that is supported by
\bref{cons}. At the same point the values of the functions are
constants
\begin{equation}
\label{condi}
r(\sigma_0) = {\rm constant}_1\,,\quad \rho(\sigma_0) = {\rm constant}_2\,,\quad
r^\prime(\sigma_0) = 0\,,\quad \rho^\prime(\sigma_0) = 0\,.
\end{equation}
We are able to find a solution for \bref{eq-r1} and \bref{eq-rho1}
under the conditions \bref{condi}. We choose the initial value of
the constants in order to get an over-critical behaviour, but
otherwise the solution is rather general. The solution fails to be
periodic in $\rho(\sigma)$ as fig. \ref{sol} shows and the
function $r(\sigma)$ behaves as expected.

As a conclusion we can say that in the case of \CN=1 SYM the
background of D5-brane, \bref{metric}, fails to give any stable
configuration in $\mathbb{R}^4\times \mathbb{R}$. Just by direct
inspection one can reach the same conclusion for configurations in
the background of the S-dual metric, \bref{S-metric} since after a
simple rescaling of the $\rho$ coordinate, this background --for
the part that is relevant here-- reduces to the flat space
considered in section \ref{secr4}. The fact that our
configurations are not stable suggest that what are we really
seeing are decaying solutions in the gravity side. If we knew
better the dictionary of the correspondence in this ${\cal N}=1$
case, we could identify the corresponding operators in the dual
gauge theory side. Then the unstability of our configurations
would signal that in strong coupling these operators quickly
branch into other operators. Since this result is beyond the
perturbative control of the gauge theory, we can conclude that the
unstability discussed above in the gravity side may encode the
information of these non-perturbative aspects of the behaviour of
certain operators in the gauge theory.

\subsection{ Poincar\'e vs. Global coordinates}

In the previous section we have showed the impossibility of
stabilising the string in  the $\mathbb{R}^4\times \mathbb{R}$
space, and thus, there is no hope to obtain a relation between the
energy and the angular momentum (or spin) close to \bref{E-S-QCD}.
In view of the field theory findings \cite{floratos} one might
wonder about the failure of the procedure leading to
\bref{E-S-QCD}.

In what follows we shall comment on the reason of this drawback.
Lets start rewriting \bref{metric} as
\begin{equation}
\label{new} ds^2 = f(\rho) \left ( dx_{0,3}^2 +d\rho^2 +
\ldots\right)\,,
\end{equation}
where ellipsis stands for compactified coordinates transverse to
the $N$ D-branes. If we compare with the $AdS_5 \times S^5$ metric
in Poincar\'e coordinates
\begin{equation}
\label{ads1} \left(ds^2\right)_{AdS_5 \times
S^5}=\frac{R^2}{\rho^2} \left(dx_{0,3}^2 + dx_p
dx_p\right)\,,\quad \rho^2=x_p x_p\,, \quad (p=4\,,\ldots\,,9)\,,
\end{equation}
we conclude that even if \bref{new} refers to a global patch (in
the sense that is not singular) its form is like written in
Poincar\'e coordinates as \bref{ads1}. As is known in the $AdS_5
\times S^5$ case, the relation \bref{E-S-QCD} can only be obtained
in global coordinates and not in Poincar\'e \cite{tseytlin}. In
the latter case is also impossible to obtain conical-like
configurations as those depicted in fig. \ref{fig}, even though in
this case it is less clear how to properly deal with the variable
$\rho$ at the origin.

\begin{figure}[t]
\begin{center}
\epsfig{file=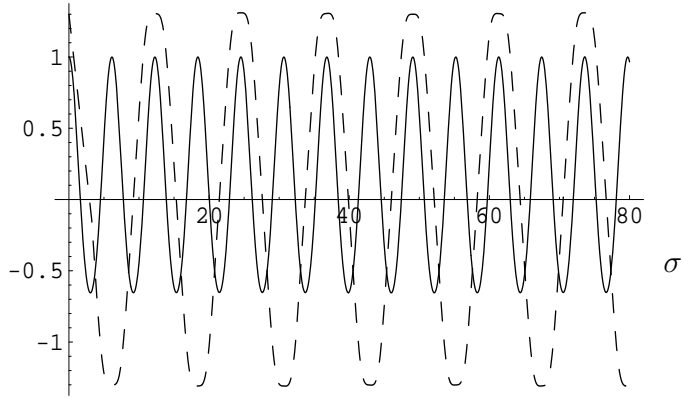,width=9cm,height=6cm}
\end{center}
\vspace{-3.0cm}\hspace{12.5cm}$\sigma$
\vspace{3.0cm}\hspace{-11.5cm} \caption{Using the ``wrong sign''
for $\Phi_D$, we get stable conical-like configurations for the
string. Here we represent the solutions $r(\sigma)$ (full line)
and $\rho(\sigma)$ (dashed curve), with some specific initial
conditions. } \label{sol2}
\end{figure}

One can conclude erroneously from the above discussion that it
might be impossible at all to find a gravity description for a
gauge theory in a non-compact $\mathbb{R}^4$ of the form
\bref{new} admitting stable configurations as those in
\bref{conflast}. Astonishingly enough, the trial $f(\rho) =
e^{-\Phi_D}$ gives a permitted solution --depicted in fig.
\ref{sol2}. It represents a string singly folded in the variable
$\rho$ and doubly folded in $r$, even though the periodicity may
depend on the exact details of the initial conditions. This ``fine
tuning'' of the initial conditions in $\sigma$ is explained by the
fact that we are looking for solutions, see \bref{conflast}, that
represent strings rotating homogeneously, that is, at constant
velocity. If the initial parameters are not correctly adjusted,
then the configuration develops oscillations in time along the $r$
and $\rho$ directions, that imply that the rotation will no longer
be homogeneous and the periodicity of the variables $r$ and
$\rho$ will be spoiled.

Also, using for $f(\rho)$ several trial functions, we are able to
find stable configurations of the same kind, or even with
different folding ratios between the $r$ and $\rho$ variables. We
think that this type of configurations, simultaneously stretched
along a parallel direction in the unwrapped part of the D-branes
and a transversal direction, may play a role in backgrounds that
allow for their stability, of which the one considered here, with
the ``wrong sign'' for the function $\Phi_D$, has been just a toy
example.

\subsection{More on \CN=1 SYM }

It is interesting to stress that although we have found
conical-like stable configurations in the preceding section, they
do not correspond to an specific background solution of string
theory. One may ask whether this can be the case for other \CN=1
SYM models as the one presented in \cite{ks}. The models \cite{mn}
and \cite{ks} describe essentially in the same manner the infrared
region but they differ substantially in the ultraviolet. This is
the reason we find \cite{ks} worthy of considering. The
construction of the supergravity dual is based on an $\SU(M+N)
\times \SU(N)$ gauge group that can be obtained by locating $N$
D3-branes and $M$ wrapped D5-branes at the tip of a conifold. In
the infra-red the theory cascades down to $\SU(M)$ if $N$ is
multiple of $M$ and the conifold is replaced by a deformed
conifold \cite{kw},\cite{kt}. Then the 10 dimensional metric is a
warped product of
 $\mathbb{R}^{3,1}$ and the deformed conifold
\begin{equation}
\label{metricks}
ds_{10}^2=h^{-1/2}(\rho) dx_{0,3}^2 + h^{1/2}(\rho) \frac{\epsilon^{4/3}}{6 K(\rho)^2} d\rho^2
+\ldots\,,
\end{equation}
where the ellipsis stand for certain tensor products of
differentials of angular variables of the Calabi-Yau metric of
the deformed conifold. We assume that via the equations of motion
we can fix the values of the angular variables in order to place
the string in a stable configuration. The functions
$K(\rho)$ and $h(\rho)$ are given by
\begin{equation}
K(\rho) = \frac{\left(\sinh(2\rho)-2\rho\right)^{1/3}}{2^{1/3}\sinh(\rho)}\,,\quad
h(\rho)= M^2 2^{2/3} \epsilon^{-8/3} \int_\rho^\infty dx\,
\frac{x \coth x -1}{\sinh^2 x} \left(\sinh(2 x)- 2 x\right)^{1/3}\,.
\end{equation}
In the sequel we shall set $\epsilon=12^{1/4}$ \cite{ks}.
As in section \ref{secr4} we consider configurations
\bref{conflast} but in the background \bref{metricks}. The equations of motion
for the radial variables are
\begin{eqnarray}
\label{r2}
r^{\prime\prime} + r \omega^2 e^2 + \frac{1}{A(\rho)}\frac{\partial A}{\partial \rho}
\rho^\prime r^\prime = 0\,, \\
\label{rho2}
\rho^{\prime\prime} + \frac{1}{2B(\rho)} \frac{\partial B}{\partial \rho} \left(\rho^\prime\right)^2
-  \frac{1}{2B(\rho)} \frac{\partial A}{\partial \rho} \left( e^2 + \left(r^\prime\right)^2
-r^2 \omega^2 e^2 \right)=0\,,
\end{eqnarray}
where we have defined
\begin{equation}
A(\rho)\equiv h^{-1/2}(\rho)\,,\quad B(\rho)\equiv \frac{\epsilon^{4/3}}{6} \frac{h^{1/2}(\rho)}
{K^2(\rho)}\,.
\end{equation}
In addition we also have the constraint
\begin{equation}
B(\rho) \left(\rho^\prime\right)^2 + A(\rho)\left(-e^2 + (r^\prime)^2 + r^2 e^2 \omega^2\right)=0\,.
\end{equation}
The solution of the system \bref{r2}-\bref{rho2} is by far more
involved that \bref{eq-r1}-\bref{eq-rho1}. There is anyhow a
simple argument to guess the solution of \bref{r2}-\bref{rho2}
with the boundary conditions \bref{condi}: as one infer from the
behaviour of the function $A(\rho)$ and $B(\rho)$ in the deep
ultraviolet
\begin{equation}
\frac{1}{A(\rho)} \frac{\partial A}{\partial \rho} \underset{(\rho
\,\, {\rm large})}{\rightarrow} {\rm constant}_1\,,\quad
\frac{1}{B(\rho)} \frac{\partial A}{\partial \rho} \underset{(\rho
\,\, {\rm large})}{\rightarrow} 0\,,\quad \frac{1}{B(\rho)}
\frac{\partial B}{\partial \rho} \underset{(\rho \,\, {\rm
large})}{\rightarrow} {\rm constant}_2\,,
\end{equation}
the system \bref{r2}-\bref{rho2} partially decouples and the
function $\rho(\sigma)$, as in the previous background, becomes
again non-periodic. We have checked numerically these
expectations. In view of the results obtained in \cite{buchel}, where from a
similar model \cite{KT} a relation like \bref{E-S-QCD} was obtained, our
conclusions are
not clear apriori. The key difference is that while the model presented in
\cite{KT} is based on $AdS$ space \cite{ks} is not assimptotic to $AdS$.

\section{Summary}
\label{summ}
\setcounter{equation}{0}

We have discussed several closed folded string configurations with
stable motion in the Maldacena-Nu\~nez supergravity background.
For strings rotating in the $S^2$, which is the cycle wrapped by
the D-branes, we find a relationship, \bref{eJ}, between the
energy and the constant $\lambda^\prime\equiv J/R$\, that is
naturally interpreted as a relationship for the energy levels of
KK stringy modes on the $S^2$. There is fairly good indication of
the decoupling of these modes in the field theory side for
operators with large values of the R-charge. {F}urthermore, in the
parallel plane-wave limit we are forced to consider a double
scaling limit in order to properly decouple the KK states. A
scaling $J \sim R$ is sufficient for this purpose. When we study
the limit of these configurations for small values of the
transversal variable $\rho$ we find that, due to the twisting that
has been introduced in order to the Maldacena-Nu\~nez background
to preserve some supersymmetry, the relation between the energy
and the R-charge, \bref{e2-short}, is not exactly that of flat
space, resulting in this case in a change of the slope for the
leading Regge trajectory.

We show also stable configurations for strings in the $S^2$ and
oscillating along the transversal direction, given by the variable
$\rho$. In this case the expression for the energy levels exactly
matches (with $\sqrt{E}$ instead of $E$) that of a N=2
super Sine-Gordon model \cite{russo-tseytlin}. This could
suggest that in the deep ultraviolet the model could define an
integrable system. We also note that our results do not belong to
the same equivalence class of the AdS models at finite
temperature.

Strings rotating in the transversal $S^3$ exhibit a behaviour
identical to that found in \cite{gkp} for strings rotating in the
$S^5$ part of the $AdS_5\times S^5$.

Looking for a non-trivial relation between the energy and the
spin, we have considered configurations stretching simultaneously
on a radial variable in the non-compact directions of the D-branes
and in the transversal variable $\rho$. We find that such
configurations are unstable in the specific background of
\cite{mn} and we only get trivial flat space rotating strings
located at $\rho=0$. We also verify that the same results
apply for the Klebanov-Strassler background. We discuss
nevertheless the possibility of having stable configurations of
this kind for other backgrounds and, surprisingly enough, we find
that a background with a change of sign of the function $\Phi_D$
present in the metric \bref{metric} allows for this type of
configurations, 
but unfortunally so far we can not interpreted from the
point of view of the supergravity solutions.
We have not neither pursue an exhaustive analysis involving compact
direction in the metric as in \cite{Hartnoll:2002th}. Perhaps one can find
a relation similar to \bref{E-S-QCD}, but is less clear how to interpret it
form a physical point of view.

\vskip 6mm

{\it{\bf Acknowledgements}}

We are grateful to Mohab Abou Zeid, Carlos N\'u\~nez, Jorge Russo
and Arkady Tseytlin for useful discussions and comments. This work
is partially supported by MCYT FPA, 2001-3598, CIRIT, GC
2001SGR-00065, and HPRN-CT-2000-00131. J.M.P. acknowledges the
Spanish ministry of education for a grant.

 \vskip 4mm

\appendix
\renewcommand{\theequation}{\Alph{section}.\arabic{equation}}
\section{Constraints in the Nambu-Goto action}
\label{app}
\setcounter{equation}{0}

A standard analysis of the Nambu-Goto (NG) action yields the
conclusion that there are no Lagrangian constraints. The reason is
as follows. Consider the NG action for the bosonic string in an
arbitrary target space background
\begin{equation}
{\cal L}_{NG} = \sqrt{| g_{\alpha \beta}|}
\end{equation}
with $g_{\alpha \beta}= G_{ij}(X)\partial_\alpha X^i
\partial_\beta X^j$ .
Under general diffeomorphism invariance,
$$\delta X^i =
\epsilon^\alpha\partial_\alpha X^i ,
$$
for an arbitrary infinitesimal diffeomorphism
$\epsilon^\alpha(\tau,\sigma)$, ${\cal L}_{NG}$ behaves as a
scalar density, that is,
\begin{equation}
\delta {\cal L}_{NG} = \partial_\alpha(\epsilon^\alpha {\cal
L}_{NG}) ,
\end{equation}
and considering that
\begin{equation}
\delta {\cal L}_{NG} = [{\cal L}_{NG}]_i \delta X^i +
\partial_\alpha({\partial{\cal L}_{NG}\over \partial (\partial_\alpha
X^i)}\delta X^i) ,
\end{equation}
(where $[{\cal L}_{NG}]_i$ stands for the Euler-Lagrange
derivative) we end up with the Noether relation
\begin{equation}
 [{\cal L}_{NG}]_i \delta X^i +
\partial_\alpha({\partial{\cal L}_{NG}\over \partial (\partial_\alpha
X^i)}\delta X^i - \epsilon^\alpha {\cal L}_{NG}) = 0
\end{equation}
identically. Out of this relation, and taking into account the
arbitrariness of $\epsilon^\alpha$, we obtain the purported
Lagrangian constraints
\begin{equation}
{\partial{\cal L}_{NG}\over \partial (\partial_\alpha
X^i)}\partial_\beta X^i - \delta^\alpha_\beta {\cal L}_{NG} = 0,
\end{equation}
which are nothing but the components of the string worldvolume
energy-momentum tensor. But it turns out that these constraints
are void, as one can check that they are mere identities
\footnote{Another way to look at this fact (no Lagrangian
constraints for the NG action) is through the canonical formalism.
One can show that there are only two primary Hamiltonian
constraints and that they are first class. In such case, it is
easy to prove that there can no be Lagrangian constraints.}.

Nevertheless, as we shall see below, special configurations for
the string may introduce constraints which are not standard in the
Dirac sense. Here we are interested in configurations describing
rotating folded closed strings, and the effect of the appearance
of a non-Dirac constraint, which is caused by the folding, is
already present in the simplest of cases, the folded closed string
rotating in flat space.

The relevant part of the metric is just
\begin{equation}
ds^2 = -dt^2 + dr^2 + r^2 d \varphi^2 ,
\end{equation}
and we consider a configuration
$$
t = \tau , \quad \varphi = \omega \tau , \quad r(\sigma) .
$$
Note that the first equality fixes the gauge for $\tau$
reparameterisation. The NG action becomes
\begin{equation}
{\cal L}_{NG} = |r^\prime|\sqrt{1-r^2 \omega^2}\,.
 \label{lng}
\end{equation}
Our configuration is intended to describe a closed string rotating
around its centre of mass, located at $r=0$, folded, and
stretching along the radial direction, with symmetric (with
respect to $r=0$) turning points. A partial use of the remaining
gauge freedom allows us to consider four pieces composing the
string, the first piece being for $\sigma \in [0, \pi/2]$ with
$r^\prime(\sigma) \geq 0$, and similar expressions for the rest. The
action then becomes
\begin{equation}
S = 4 \int d\tau \int_0^{\pi/2}  d\sigma r^\prime(\sigma) \sqrt{1-r^2
\omega^2} . \label{sng}
\end{equation}
The only variable left, $r(\sigma)$, makes the Lagrangian to be a
total derivative -with respect to the $\sigma$ coordinate.
Therefore, at first sight, it would seem as if the function
$r(\sigma)$ remains completely arbitrary, for there will be no
e.o.m. for it, except for the only natural requirement of the
positivity of the term in the square root, that is, $r \leq
1/\omega$. But this analysis is incomplete, as we shall show now,
because there is a boundary effect, caused by the folding of the
string, that has been overlooked.

For the first quarter of the string, with parameterisation $\sigma
\in [0, \pi/2]$, the string will stretch from $r=0$ to some $r=r_0
\leq 1/\omega$, the action can then be written
\begin{equation}
S = 4 \int d\tau \int_0^{r_0}  dr \sqrt{1-r^2 \omega^2} .
\label{sng2}
\end{equation}
Now the action for this configuration of the string is determined
 by a single parameter $r_0$. The correct
application of the variational principle requires the action to be
extremised with respect to this parameter, that is,
\begin{equation}
{\partial \over \partial r_0 } \int_0^{r_0}  dr \sqrt{1-r^2 \omega^2}\,,
\end{equation}
giving
\begin{equation}
\sqrt{1-r_0^2 \omega^2} =0\,,
\end{equation}
implying,
\begin{equation}
r_0 = {1  \over \omega}\,.
\end{equation}
This is the constraint that restricts the setting of initial
conditions for the string. It is not of the usual Dirac type, for
it only restricts the
 positions of two points -the turning points- of the whole string.
 It is a constraint
that tells how far in the radial direction the
 string stretches. It turns out that it stretches all that it can:
 until
 the extremes move at the speed of light.

 This result is obtainable in
 a straightforward way in the conformal
 gauge, but it is easy to be missed using the NG approach (in fact,
 as we have observed, since the Lagrangian
(\ref{lng}) is
 locally a total derivative -in $\sigma$-  one could naively
 expect that there will no be e.o.m. for $r$, and so the speed of light
condition will be overlooked). What we have shown, therefore, is the
 equivalence between the NG and the conformal gauge approaches for the
 description of this type of configurations. This is something
 that is
 formally guaranteed by the equivalence of gauges, but that does not
 spares us of the subtleties involved. Observe in addition that
 when one works in the conformal gauge, the conformal factor becomes
 singular at the turning points.

It is worth noting
 that this type of configurations of folded closed strings has the same
 dynamics as open strings with Neumann boundary conditions.

\end{document}